\documentclass[aps,prb,showpacs,twocolumn]{revtex4}  

\usepackage{graphicx}
\usepackage{amssymb}                

\def\newblock{\hskip .11em plus .33em minus .07em}
\def\gapx{\lower 2pt \hbox{$\buildrel>\over{\scriptstyle{\sim}}$\ }}
\def\lapx{\lower 2pt \hbox{$\buildrel<\over{\scriptstyle{\sim}}$\ }}
\def\he4{$^4$He}
\def\paraH2{{\it p}-H$_2$}
\def\orthoD2{{\it o}-D$_2$}
\def\Am2{\AA$^{-2}$}

\begin{document}

\title{Systematics of small parahydrogen clusters in two dimensions}
\author{Saheed Idowu and Massimo Boninsegni} 
\affiliation{Department of Physics, University of Alberta, Edmonton, 
    Alberta, Canada T6G 2E7}
\date{\today}

\begin{abstract} 

We studied by means of  computer simulations the low temperature  properties of two-dimensional parahydrogen clusters comprising between $N=7$ and 30 molecules.
Computed energetics is  in quantitative agreement with that reported in the only previous study [Phys. Rev. B {\bf 65}, 174527 (2002)], but a generally stronger  superfluid response is obtained here for clusters with more than ten molecules. Moreover, all the clusters, including the smallest one,  display a well-defined, clearly identifiable solidlike structure; with only one possible exception, those with fewer than  $N=25$ molecules are  (almost) entirely superfluid at the lowest temperature considered here (i.e., $T$=0.25 K), and can thus be regarded as nanoscale ``supersolids". The implications of these results on a possible bulk  two-dimensional superfluid phase of parahydrogen are discussed.

\end{abstract}

\maketitle

\section{INTRODUCTION}
The physics  of small parahydrogen clusters is of interest because these few-body systems feature a rather unique interplay of classical and quantum physics.\cite{holland} Due to its low mass and bosonic character, parahydrogen (\paraH2) was predicted a long time ago to undergo a superfluid transition, at temperature $T \lesssim$ 6 K.\cite{ginzburg72} However, the equilibrium phase of bulk \paraH2 in the $T\to 0$ limit is a crystal, even in reduced dimensions,\cite{boninsegni04,boninsegni13} due to the depth of the attractive well of the interaction potential between  two hydrogen molecules. The theoretical suggestion that a liquidlike phase of \paraH2 could be stabilized in two dimensions (2D) by an underlying impurity substrate\cite{gordillo97} is not supported by several subsequent calculations.\cite{boninsegni05,turnbull}
\\ \indent
There is experimental evidence, on the other hand, that small droplets of \paraH2  in three dimensions can escape crystallization,\cite{prozument} down to a temperature sufficiently low ($\sim$ 1 K) that a finite superfluid response, defined as the dissipationless rotation about an axis going through the center of mass,  is theoretically predicted to arise in nanoscale clusters ($\lesssim 30$ molecules).\cite{sindzingre,noi,noi2}. Most of these clusters are expected to remain  essentially liquidlike, i.e., structureless, all the way to zero temperature, in some cases undergoing quantum melting at sufficiently low $T$; one of them, however, namely (\paraH2)$_{26}$, is predicted to retain some well-defined structural short-range order, even with the concurrent development of superfluid coherence at low $T$, behaving in some sense as  a finite-size ``supersolid".\cite{toennies}
\\ \indent
An interesting question is what happens if clusters are themselves confined to 2D, which could be  experimentally achieved for example by adsorbing a low-density \paraH2 film on a suitable substrate, strong enough to confine molecules effectively to 2D, but also weak enough to allow for the neglect of corrugation. Substrates of alkali metals might be a good candidate,\cite{chy} but progress toward the stabilization of quasi-2D H$_2$ clusters on a different type of substrate has been recently reported.\cite{nano}
Reduction of dimensionality brings about two competing effects. On the one hand, the lower coordination number weakens the tendency to crystallize, but also has the effect of rendering less frequent quantum-mechanical exchanges of identical particles, which have been shown to play a crucial role in the stabilization of a liquidlike structure at low $T$.\cite{noi,noi2}
\\ \indent
The only existing theoretical study of \paraH2 clusters in 2D is that by Gordillo and Ceperley, who carried out first principles Path Integral Monte Carlo simulations, down to $T=0.3$ K.\cite {Ceperley02} Their main physical findings were that clusters comprising at the most two concentric shells of molecules around the center of mass (typically ten molecules or less) are  liquidlike and superfluid at low $T$ ($\lesssim 1$ K); as more molecules are added to the cluster, a solidlike core starts forming, with the concomitant, gradual suppression of the superfluid response. From this observation, the authors inferred that, as cluster size is increased, superfluidity is progressively confined to a liquidlike outer shell, while the solidlike core is insulating.
\\ \indent
In this paper, we present results  of Quantum Monte Carlo simulations of  2D \paraH2 clusters of size ranging from $N=7$ to $N=30$ molecules at low temperature (down to 0.25 K).  We computed energetic and superfluid properties of the clusters, and investigated their structure by means of density profiles computed with respect to the center of mass of the cluster, as well as through actual density maps, affording direct visual insight in 2D.\\
Our estimates of the energy per \paraH2 molecule are in excellent quantitative agreement with those of Ref. \onlinecite{Ceperley02}; on the other hand, we obtain a stronger superfluid signal than they do, especially for clusters with more than $\sim 10$ molecules. Indeed, we find a robust superfluid signal at the lowest temperature considered here, for clusters comprising as many as 
$N=25$ \paraH2 molecules. Analogously to what observed in three-dimensional (3D) clusters, the dependence of the superfluid response on $N$ is non-monotonic.\cite{noi,noi2}
\\ \indent
More importantly, the physical picture that emerges from our study is qualitatively different from that of Ref. \onlinecite {Ceperley02} in a number of relevant aspects.
First and foremost, none of the clusters studied here can be regarded as truly ``liquidlike", at low $T$. In particular, our radial density profiles for the smaller clusters are quantitatively very different from those of Ref. \onlinecite{Ceperley02}, 
featuring much higher peaks, separated in turn by much more pronounced dips. This is indicative of a well-defined structure, in which molecules tend to occupy preferred lattice positions, something that is confirmed by our computed density maps.
Second, the calculation of the local superfluid density shows that in all superfluid clusters the response is not confined at the surface but rather uniform throughout the system, much like in 3D clusters.\cite{noi3} Thus, no meaningful distinction can be drawn between a non-superlfuid, solidlike center, and a superfluid liquidlike outer part, for any of the superfluid clusters; rather, they should be regarded as featuring concurrently superfluid and solidlike properties. In this sense, these small 2D clusters may be regarded as naturally occurring nanoscale ``supersolids".
\\ \indent
The remainder of this article is organized as follows: in the next section we describe the microscopic model underlying the calculation and  furnish basic computational details; we devote Sec. \ref{results} to a thorough illustration of our results,  discussing the physical conclusions in Sec. \ref{conclusions}, in which we also describe some possible scenarios for the stabilization of a bulk superfluid phase of \paraH2 in 2D.

\section{MODEL AND METHODOLOGY}
\label{model}

Our system of interest is modeled as a collection of $N$ {para}hydrogen (\paraH2) molecules, regarded as point particles, moving in two physical dimensions. The quantum mechanical many-body Hamiltonian is the same as in Ref. \onlinecite{Ceperley02}, given by:

\begin{eqnarray}\label{hm}
\hat{H}&=&-\lambda\sum_{i=1}^{N}\nabla_{i}^{2}  + V(R)
\end{eqnarray}
where $\lambda= 12.031$ K \AA$^{2}$, $R\equiv {\bf r}_1,{\bf r}_2, ...{\bf r}_N$ is a collective coordinate referring to all $N$ particles in the system and $V(R)$ is the total potential energy of the configuration $R$, which is assumed here to be expressed as the sum of pairwise contributions, each one described by a spherically symmetric potential. In this calculation, we made use of Silvera-Goldman potential,\citep{Silvera1} mostly for consistency with the  calculation\cite{notex} of Ref. \onlinecite{Ceperley02}. We computed equilibrium thermodynamic properties 
of this finite system at low temperature, by means of Quantum Monte Carlo simulations, based on the Worm Algorithm in the continuous-space path integral representation. Because this methodology is extensively described elsewhere,\cite{worm,worm2}  we do not review it here, but limit ourselves to listing a few important computational details. 
\\ \indent
Our simulated system is enclosed in a square cell, chosen sufficiently large to remove any effect of the boundary conditions, periodic in all directions. No artificial confining potential was used in the simulation, as  clusters stay together simply as a result of the intermolecular pairwise attraction.
We use a high-temperature approximation for the imaginary time propagator accurate to fourth order\cite{chin,jiang,cuervo} in the imaginary time step $\tau$. The results shown here are obtained with a value of $\tau=10^{-3}$ K$^{-1}$, empirically found to yield estimates indistinguishable, within our quoted statistical uncertainties, from those extrapolated to the $\tau \rightarrow0$ limit (i.e the limit where the method becomes exact). 
\\ \indent
We compute both the global superfluid fraction $\rho_S$, as well as the radial, angularly averaged one, $\rho_S(r)$. We estimate the first using the well-known ``area" estimator, \cite{sindzingre89} the second by means of a straightforward generalization of the area estimator, applied to concentric  shells of varying radii, centered at the center of mass of the cluster.\cite{paesani} 
For the smallest clusters ($N\lesssim 10$), a more accurate estimate of the global superfluid fraction is given by the radial average of $\rho_S(r)$, weighted by the angularly averaged radial \paraH2 density $\rho(r)$, outside of a circle of radius $r_\circ\sim$ 2 \AA. This is because the statistical noise in the estimate of $\rho_S$ arises mostly from contributions  in the vicinity (i.e., within a distance $r_\circ$ or less) of the center of mass.

\section{RESULTS}  
\label{results}
As mentioned above, we carried out numerical simulation of clusters at temperatures as low as 0.25 K. In general, structural and energetic properties of the clusters remain essentially unchanged below $T\sim 0.5$ K; in particular, physical estimates reported here for $T=0.33$ K or lower, should be regarded as ground state estimates, within their statistical uncertainties.

\begin{figure}             %               ENERGYPLOT
\centerline{\includegraphics[scale=0.7]{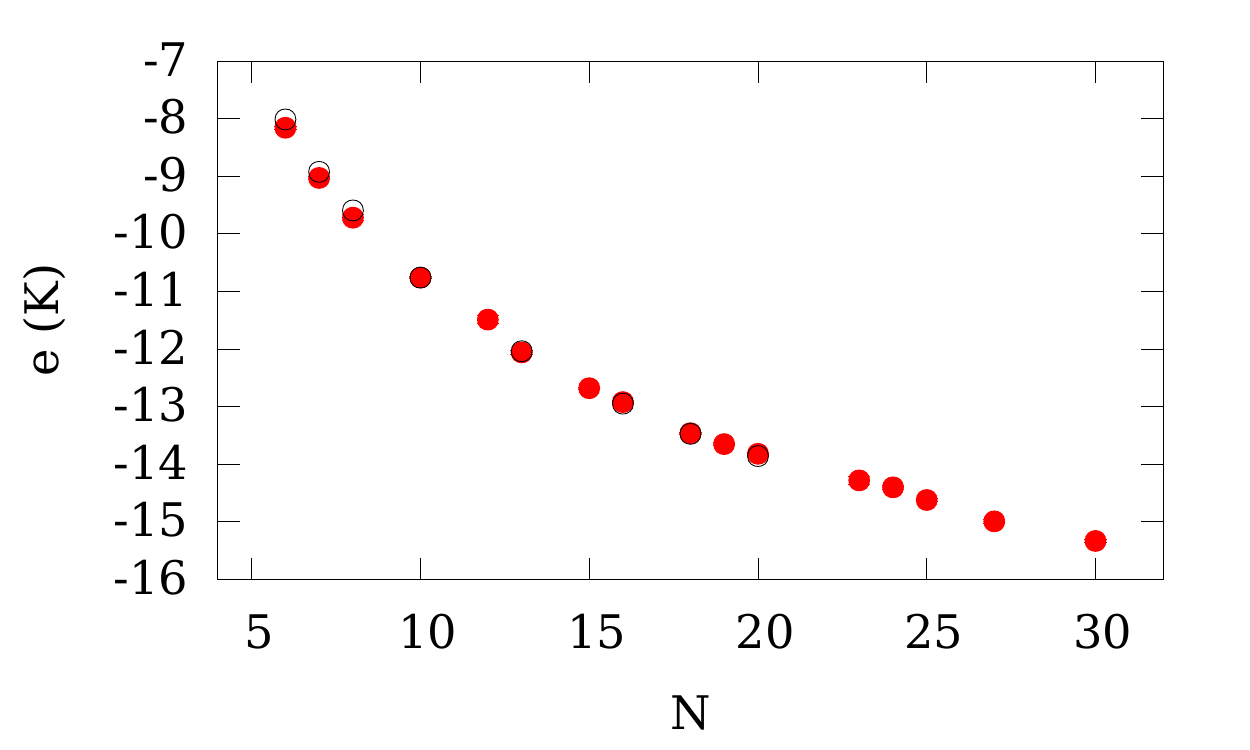}} 
\caption{(color online) Energy per hydrogen molecule $e$ (in K) versus cluster size $N$, at $T=0.25$ K (full symbols). Also shown are the results reported in Ref. \onlinecite{Ceperley02},  at $T$=0.33 K (open symbols). For clusters of size $N$=13,16 and 20 our energy estimates are indistinguishable from those of Ref. \onlinecite{Ceperley02}. Statistical errors are at the most equal to symbol size.}
\label{ene}
\end{figure} 
Fig. \ref {ene} shows the energy per molecule versus cluster size, computed at $T=0.25$ K. Also shown are the results reported in Ref. \onlinecite{Ceperley02} at a slightly higher  temperature ($T=0.33$ K). The two calculations are in excellent agreement, within their statistical uncertainties. The energy per molecule is monotonically decreasing, with no evidence of ``magic numbers", within the precision of our calculation. It attains a value around $-15.3$ K for a cluster with $N=30$ \paraH2 molecules; this is still relatively far from the 2D bulk value\cite{boninsegni04} of $\sim -23.2$ K.
\begin{figure} [h]            
\centerline{\includegraphics[scale=0.38]{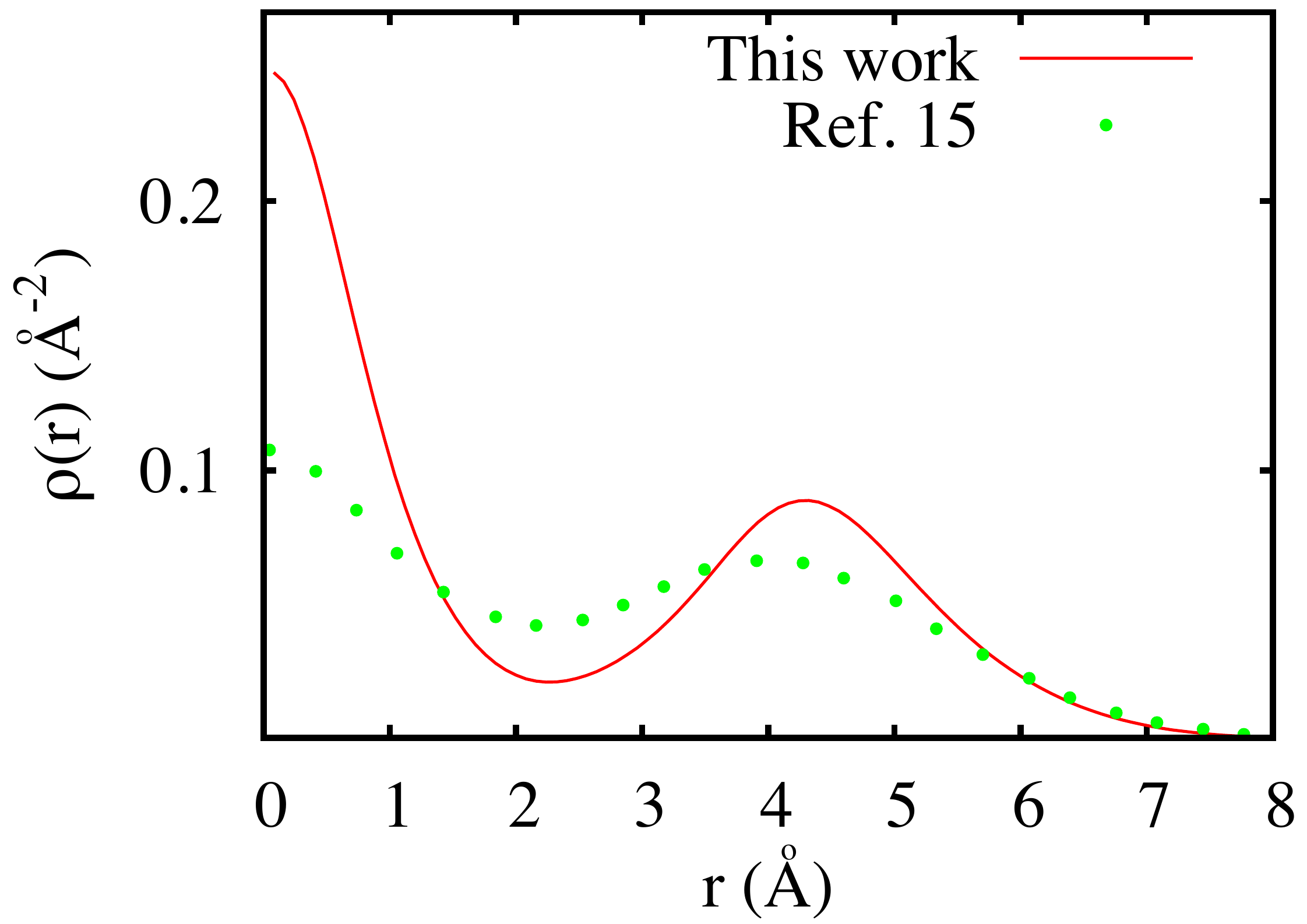}} 
\caption{(color online) Radial density profile for a cluster with $N=7$ \paraH2 molecules at $T=0.33$ K (solid line). Filled circles show the corresponding result from Ref. \onlinecite{Ceperley02}. Profiles are computed with respect to the  center of mass of the cluster. Statistical errors are not visible on the scale of the figure. The local superfluid density profile for this cluster is indistinguishable from that of the local density.}
\label{r7}
\end{figure}
\begin{figure} [h]            
\centerline{\includegraphics[scale=0.50]{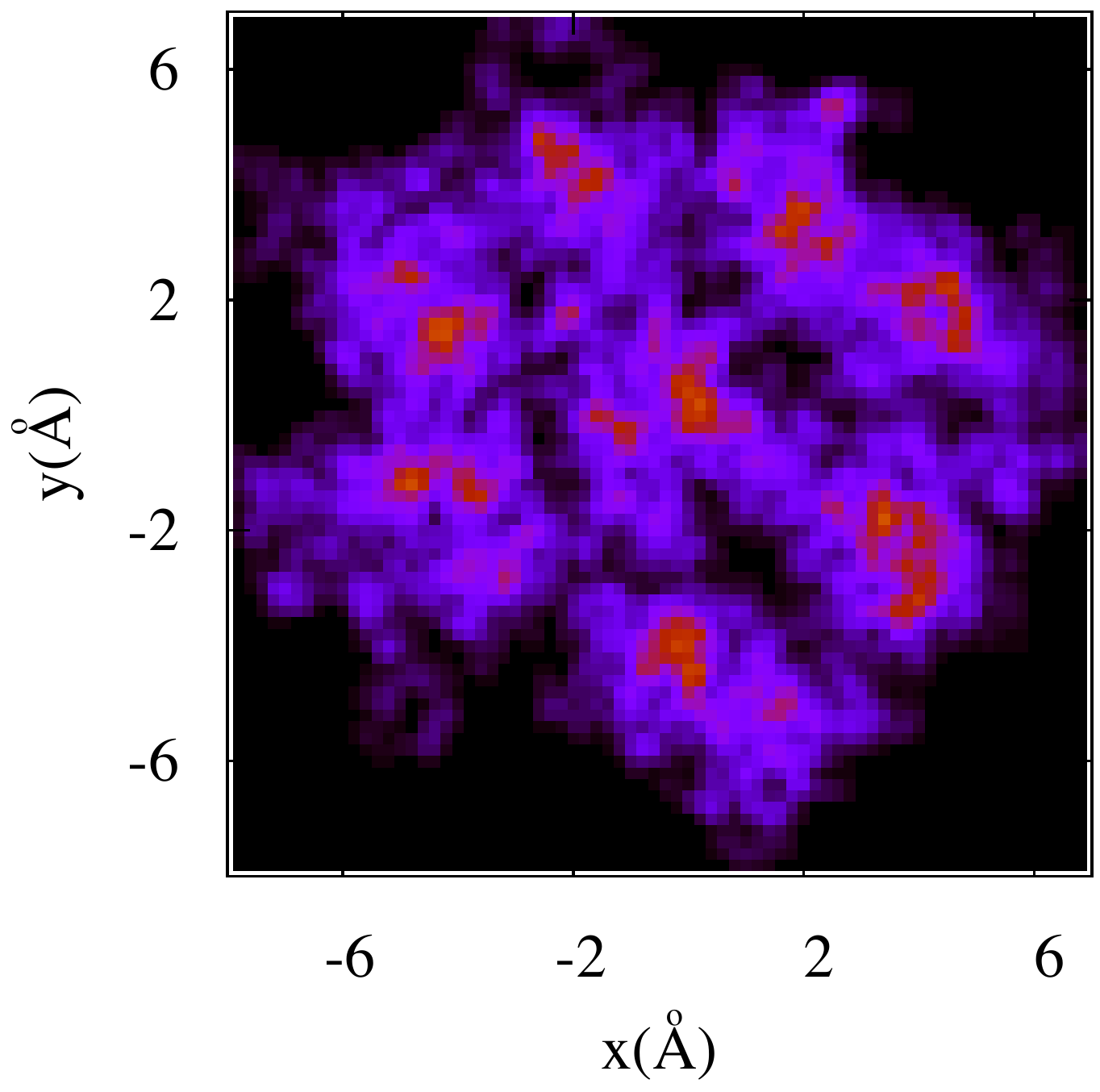}}
\caption{(color online) Configurational snapshot (particle world lines) yielded by a simulation of a cluster with $N=7$ \paraH2 molecules at $T=0.33$ K.  Brighter colors correspond to a higher local density.}
\label{sn1}
\end{figure}
\\ \indent
In order to discuss the structural properties of the clusters, which are the main focus of this study, 
we begin by illustrating in some details the results for the smallest 2D droplet studied  in this work, namely that with $N$=7 \paraH2 molecules, because in many respects this allows us to draw general conclusions, applicable to clusters of greater size as well. 
\\ \indent
Figure \ref{r7} shows the radial density profile $\rho(r)$ for  (\paraH2)$_7$  (solid line), with respect to its center of mass, computed at a temperature $T$=0.33 K. Two results are shown, namely that obtained in this work (solid line), and that published in Ref. \onlinecite{Ceperley02} (filled circles); for comparison purposes,  we begin by discussing the latter first. It displays two broad peaks, one at the origin,  signaling a particle in the center of the cluster, and an outer one, ostensibly the signature of a floppy surrounding ring comprising the remaining six molecules.  Only a minor depression between the two peaks is observed; indeed, the outer peak is barely detectable. Such a profile was reasonably interpreted by the authors of Ref. \onlinecite {Ceperley02} as evidence of a structureless, liquidlike cluster. 
\\ \indent
The corresponding density profile obtained in this work, on the other hand, looks markedly different. It  features two much higher and narrower peaks (the one at the origin more than twice as high with respect to that of Ref. \onlinecite{Ceperley02}), separated by a pronounced dip, to suggest a rather sharp physical demarcation between the central particle and the surrounding ring. This points to a considerably more structured cluster, in which particles preferentially tend to be at well-defined relative positions, i.e., the cluster is solidlike.\cite{notesame}
The quantitative disagreement between the two results is puzzling, considering that our estimate of the energy per particle ($-8.92(3)$ K), as well as that of the superfluid fraction $\rho_S$ ($\sim$ 100\%) at this temperature are in agreement with theirs, within the statistical uncertainties of the calculations. In order to obtain an independent check of our result, we carried out a separate calculation of the ground state properties of this cluster, using the Path Integral Ground State technique.\cite{cuervo,sarsa} The details of this calculation are identical with those described in Ref. \onlinecite{saverio}. The radial density profile obtained in this second way falls right on top of  that at $T=0.33$ K, given by the solid line in Fig. \ref{r7}. This fact  gives us confidence on the correctness of our results. The qualitative disagreement with the radial density profile of Ref. \onlinecite{Ceperley02} is therefore unclear, as the same microscopic model is utilized.
\\ \indent
In order to gain additional insight into the physics of this few-body system, we make use of the direct and visually suggestive information  provided in 2D by configurational snapshots generated 
by the Monte Carlo simulation. (see, for instance, Ref. \onlinecite{mio})  Fig. \ref{sn1} shows a particle density map obtained from a statistically representative  configuration snapshot (i.e., particle world lines) for the cluster under exam, at the same temperature as in Fig \ref{r7}. By ``statistically" representative, it is meant here that every configuration generated in the simulation is roughly similar to that shown in the figure,  differing from it mostly by a mere rotation.
Albeit smeared by zero point motion, lumps associated to individual molecules are clearly identifiable, forming an ordered structure, with a visible gap between the particle in middle of the cluster and those in the outer ring. In spite of this relatively ``ordered" arrangement, in turn implying a degree of molecular localization, exchanges of indistinguishable particles occur frequently, hence the large superfluid response. 
\\ \indent
The superfluid fraction of this cluster is, as mentioned above, 100\% at $T\le 0.33$ K, within  statistical uncertainty. Moreover, the superfluid signal is uniformly distributed throughout the whole cluster, not concentrated at any specific region (e.g., the surface); in fact, the computed angularly averaged, local superfluid density profile is indistinguishable from that of the local density, shown in Fig. \ref{r7}. This is much like already observed in 3D clusters.\cite{noi3} Thus, the two seemingly exclusive superfluid and solidlike properties appear to merge  into a single, remarkable ``supersolid phase".\cite{notesame}
\\ \indent
\begin{figure}             %               ENERGYPLOT
\centerline{\includegraphics[scale=0.36]{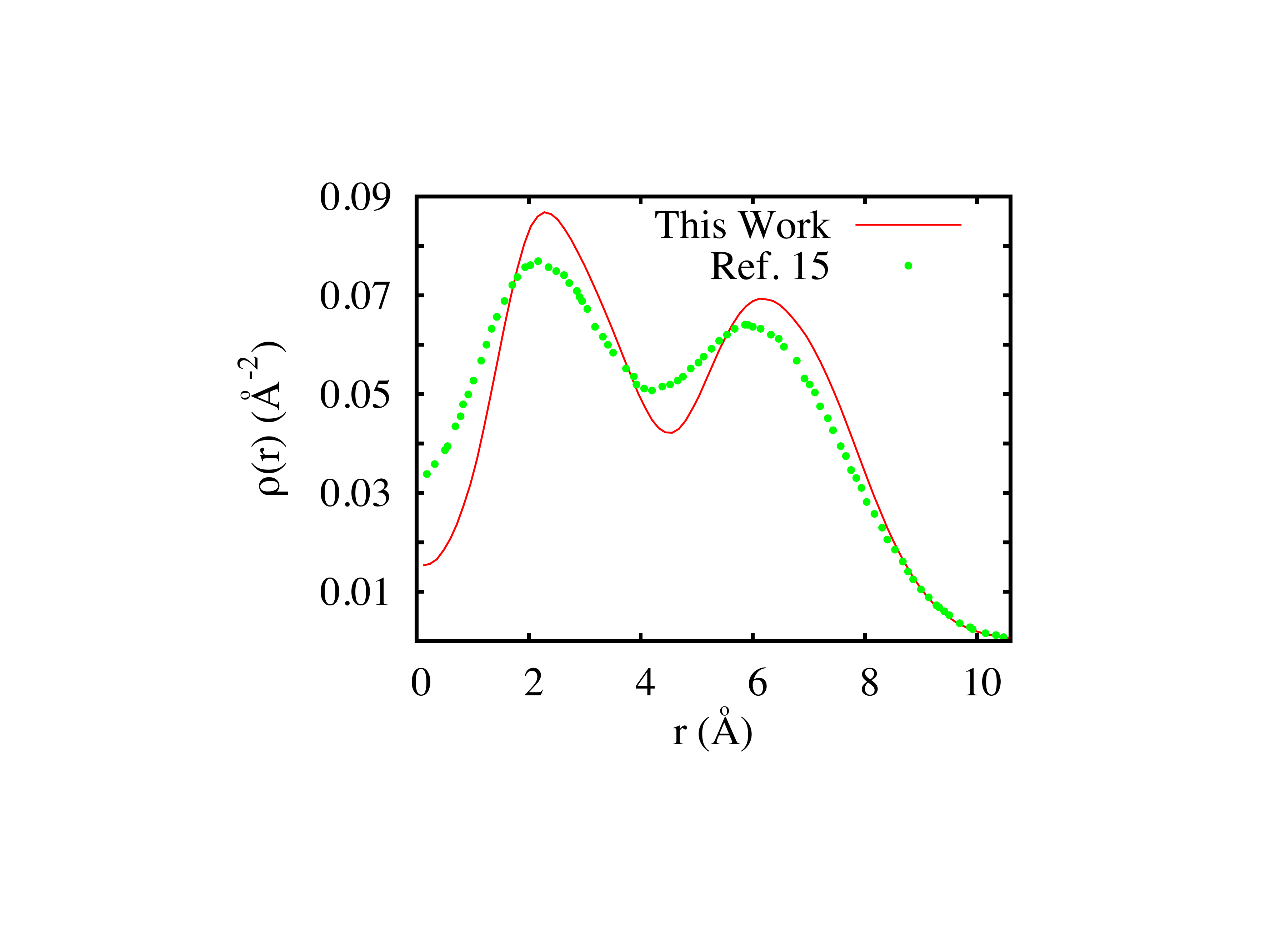}} 
\caption{(color online) Radial density profile for cluster with $N=13$ at $T=0.33$ K (solid  line). Also shown is the 
corresponding result from Ref. \onlinecite{Ceperley02} (filled  circles). Profiles are computed with respect to the  center of mass of the cluster. Statistical errors are not visible on the scale of the figure. }
\label{r13}
\vskip 0.4 in
\centerline{\includegraphics[scale=0.38]{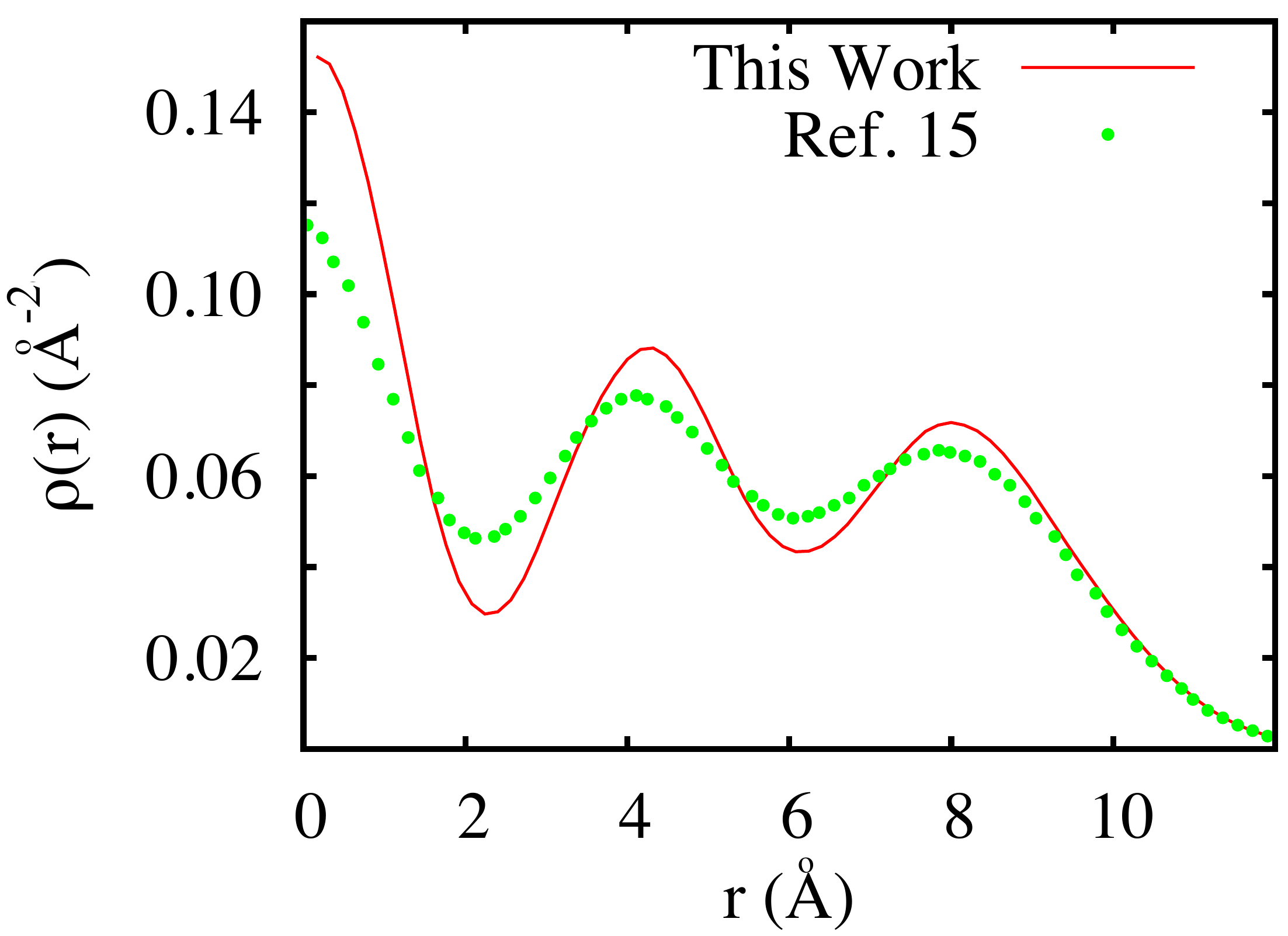}} 
\caption{(color online) Radial density profile for cluster with $N=20$ at $T=0.33$ K (solid  line). Also shown is the 
corresponding result from Ref. \onlinecite{Ceperley02} (filled circles). Profiles are computed with respect to the  center of mass of the cluster. Statistical errors are not visible on the scale of the figure. }
\label{r20}
\end{figure}
The same structural short-range order characterizing this small cluster is found in  {\it all} other clusters investigated here. All of them are solidlike, in no case quantum-mechanical exchanges causing the melting at low $T$ into a featureless, liquidlike cluster, an effect which is instead observed in simulations of 3D clusters, in this temperature range.\cite{noi,noi2} 
Figures \ref{r13} and \ref{r20} show  radial density profiles for clusters with $N=13 $ and 20 molecules, at $T$=0.33 K. Comparison with Ref. \onlinecite{Ceperley02} again shows considerably more structure in the
simulations carried out here. As the size is increased, both studies yield evidence of greater structural short-range
order, but quantitative differences between radial density profiles remain visible even for the largest cluster studied in Ref. \onlinecite{Ceperley02}, i.e., that with $N$=20 (Fig. \ref{r20}). While a  quantitative characterization of  particle localization in these clusters might be obtained by making use of estimators proposed in Refs. \onlinecite{cha,cuervo2,guardiola}, nonetheless our results, consistently yielding higher peaks and more pronounced dips inbetween, seem to point rather clearly to a  significantly more rigid, solidlike structure than that predicted in Ref. \onlinecite{Ceperley02}. The disagreement is not only  quantitative, for the smallest clusters it is even {\it qualititative}.
\\ \indent
We now discuss the superfluid properties.
For all clusters with $N\le 22$, 
the superfluid fraction $\rho_{s}$ is indistinguishable from 100\% within statistical uncertainties at $T = 0.25$ K. Our superfluid signal is stronger than that reported in Ref. \onlinecite{Ceperley02} for clusters with $N=13,\ 16$ and 20, for all of which they find
is worth $0.6\pm0.1$ at $T$=0.33 K, whereas for these four specific clusters we find values in excess of 90\% at that same temperature.  The largest cluster for which a significantly large superfluid response is observed at the lowest temperature considered here, namely $T$=0.25 K, comprises 25 molecules; its superfluid fraction is again worth approximately 100\% at $T$=025 K.
\\ \indent
The superfluid response is observed to drop abruptly for $N>25$; indeed, for none of the clusters with $26\le N\le 30$ could we obtain an appreciable value of $\rho_S$ in this study,  at the lowest temperature considered here. It should be noted, however, that the dependence of $\rho_S$ on $N$ for a fixed low $T$ is not  monotonic, a fact already observed in 3D clusters\cite{noi,noi2} with $22\lesssim N\lesssim 30$. Specifically, at $T$=0.25 K, $\rho_S$ is observed to drop down to level of statistical noise\cite{note2} for a cluster with $N$=24, to rebound to $\sim 100\%$ on adding a single molecule (i.e., for a cluster with $N$=25), and to drop again to zero if another molecule is added.
\begin{figure} [h]            
\centerline{\includegraphics[scale=0.38]{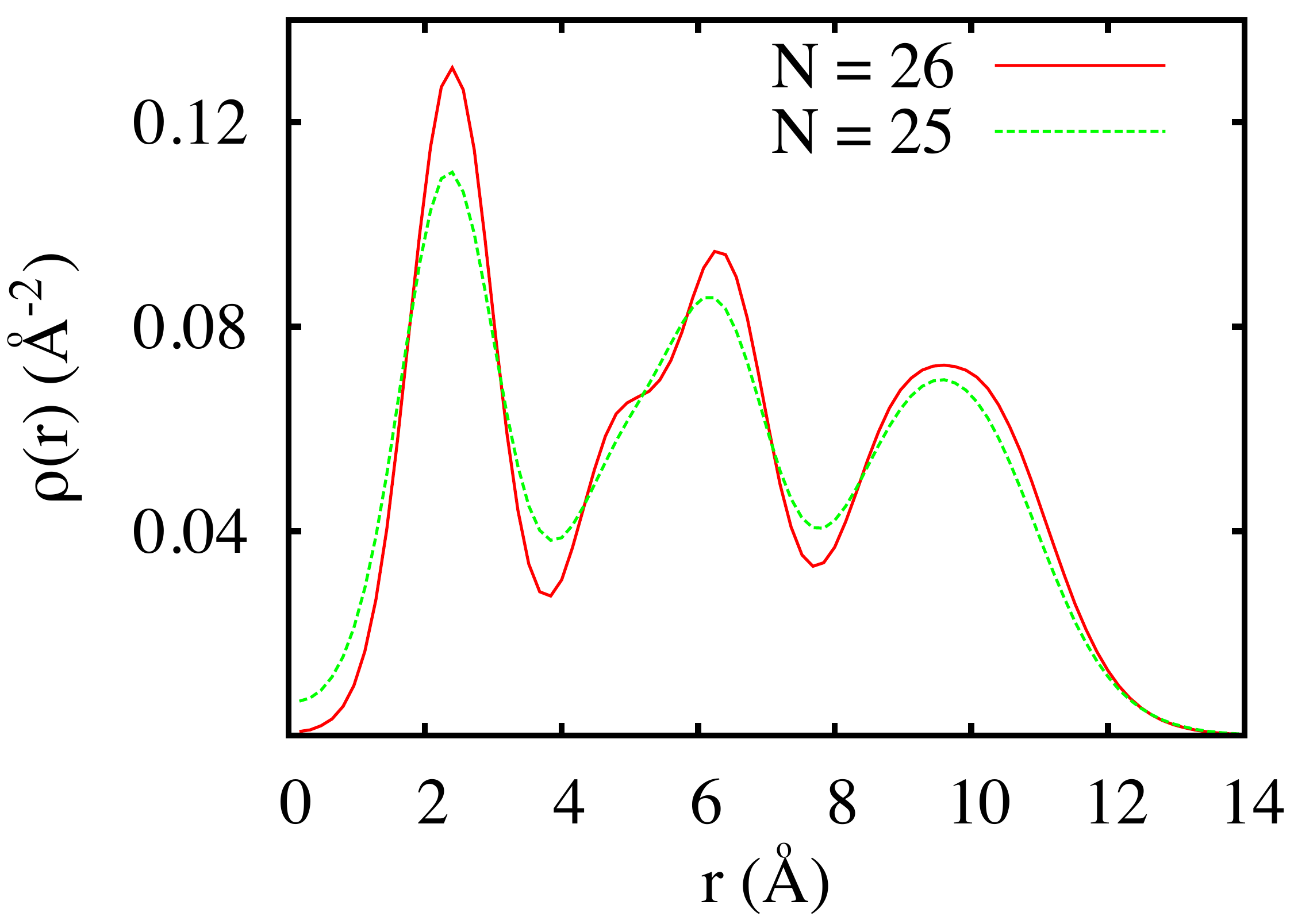}} 
\caption{(color online) Radial density profiles for clusters with $N=25$ (dashed line) and 26 (solid line) \paraH2 molecules at $T=0.25$ K. Profiles are computed with respect to the  center of mass of the cluster. Statistical errors are not visible on the scale of the figure.}
\label{r256}
\end{figure}

\begin{figure}             %               ENERGYPLOT
\centerline{\includegraphics[scale=0.50]{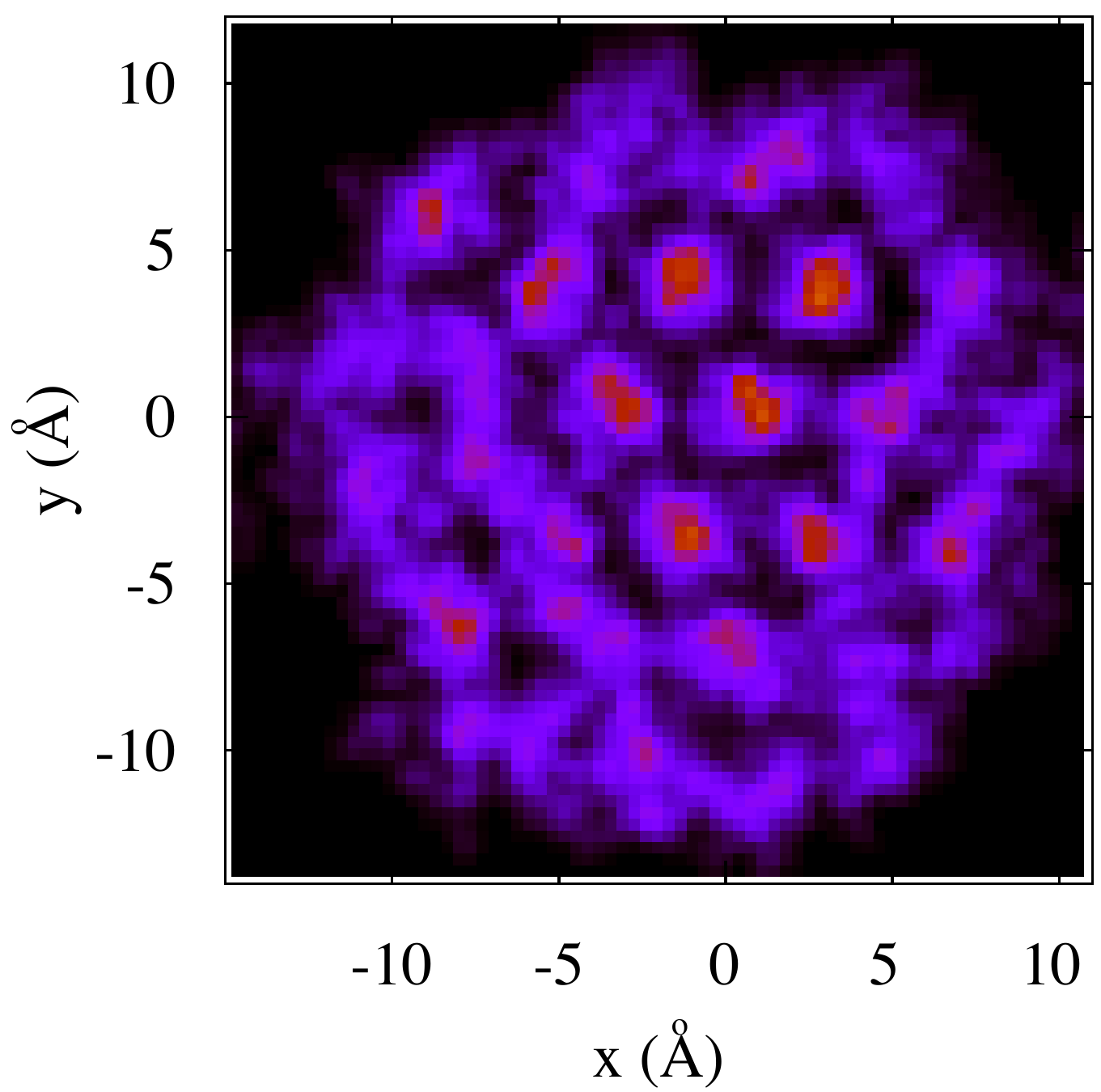}}
\caption{(color online) Configurational snapshot (particle world lines) yielded by a simulation of a cluster with $N=25$ \paraH2 molecules at $T=0.25$ K.  Brighter colors correspond to a higher local density.}
\label{sn2}
\vskip 0.4 in
\centerline{\includegraphics[scale=0.50]{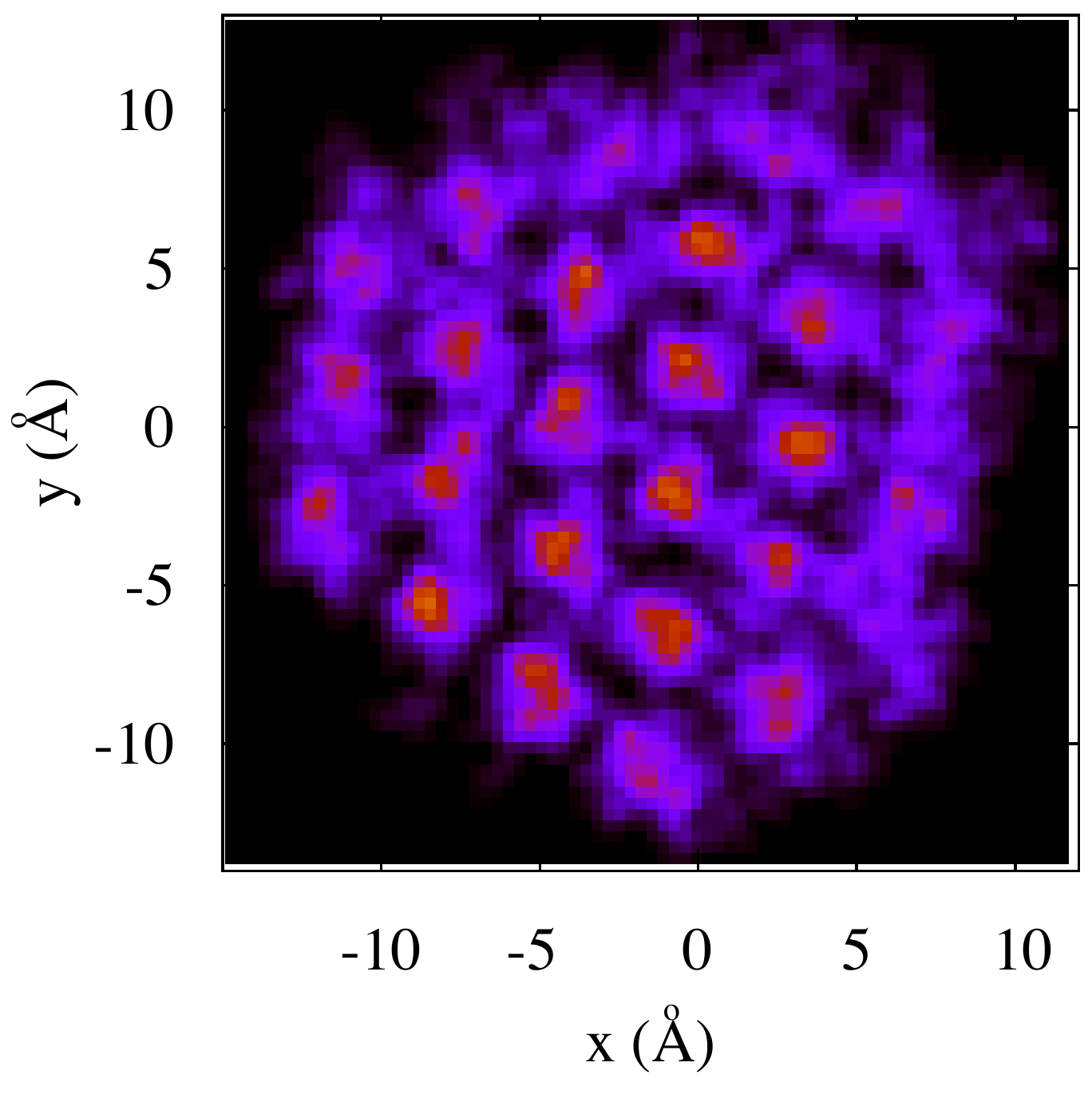}} 
\caption{(color online) Configurational snapshot (particle world lines) yielded by a simulation of a cluster with $N=26$ \paraH2 molecules at $T=0.25$ K.  Brighter colors correspond to a higher local density.}
\label{sn3}
\end{figure}
Such an intriguing behaviour, most remarkably mimicking what observed in 3D clusters with the same numbers of molecules, cannot be straightforwardly related to the shape of the cluster, and/or to the completion of any regular geometrical structure, occurring on adding a molecule to (or removing one from) a cluster with $N$=25, nor to some greater ``liquidlike" character of the $N=25$ system. Fig. \ref{r256} compares density profiles for the two clusters with $N$=25 and 26 \paraH2 molecules at $T$=0.25 K, at which the cluster with 25 molecules is entirely
superfluid, and that with 26 features no measurable superfluid response. Both profiles show three peaks, i.e., three concentric shells, but that of the the cluster with one extra molecule is noticeably more structured, and its first two
peaks sharper. On the other hand, the peak corresponding to the  outer shell  is rather smooth, very similar in both cases. Thus, the main structural difference between the two clusters seems to be that the non-superfluid one has enhanced solid order in the two inner shells. This observation, together with the fact that  the superfluid response is uniformly distributed throughout the cluster,  in turn undermines the suggestion that superfluidity should correlate with the presence of a liquidlike outer shell, of which no evidence is shown by the representative configuration snapshots of Figures \ref{sn2} and \ref{sn3}. Indeed, both  clusters display a rather ordered structure, although the molecules sitting at the surface are obviously less bound.

\section{Discussion}
\label{conclusions}
We have carried out a systematic investigation of the low temperature properties of small clusters of \paraH2 in two dimensions, using first principle quantum Monte Carlo simulations, whose only input is the intermolecular pair potential. Some of the physical properties of these clusters are very similar to those of clusters in three dimensions. For example, the non-monotonic dependence of the superfluid response at low temperature on the number $N$ of molecules in the cluster, is also observed in the 3D system. There too, most notably, the numbers $N$=25 and 26 are associated with the same effect observed here, i.e., the drop of the superfluid fraction from nearly 100\% to a value close to zero (in the temperature range considered here), on adding a single molecule to the $N$=25 cluster.  This suggests that a remarkable compensating effect takes place on reducing dimensionality; the enhancement of quantum fluctuations, and the concomitant suppression of quantum-mechanical exchanges (due to the confinement of molecular motion to a plane, and the hard, repulsive core of the intermolecular potential at short distances) both contribute to preserve some of the same physics observed in three dimensions.
\\
\indent
On the other hand, some of the features of 2D \paraH2 droplets set them aside from their 3D counterpart. 
The most striking feature of these few-body systems, is the simultaneous presence of what can be reasonably described as short-range order, whereby molecules tend to form specific geometrical arrangements, typically resulting from a classical mechanism (i.e., minimization of potential energy), and a finite superfluid response, originating from exchanges of identical molecules. Much like in the 3D case, superfluidity is underlain by cycles of exchanges involving all of the molecules, not just those on the outer shell. Indeed, the participation of inner molecules to exchanges is crucial, witness the fact that as the number $N$ is increased beyond 25, at which point the superfluid response of the clusters is significantly suppressed, their inner structure concurrently appears much more rigid (as shown by the comparison in
Fig. \ref{r256}), consistently with inner molecules to be more localized and less involved in exchanges.
 \\ \indent
What are the implications of this study, if any, regarding a possible stabilization of a superfluid phase of bulk \paraH2? The suggestion that one might be able to use frustration, either arising from disorder\cite{turnbull,long} or from an underlying impurity substrate incommensurate with the equilibrium triangular crystalline phase of \paraH2 in two dimension,\cite{gordillo97} has not been shown to lead to a superfluid phase, although recently renewed claims to that effect have been made.\cite{caz}
\\ \indent
An alternative approach could borrow on ideas arising from theoretical studies of superfluidity (and supersolidity) in cold atom assemblies.\cite{myreview} Specifically, one could imagine patterning a suitably chosen surface with regularly arranged adsorption sites (e.g., on a triangular lattice). Each site could be designed to accommodate a number of \paraH2 molecules between, say, ten and twenty, acting in a sense as a ``molecular quantum dot", turning superfluid at low $T$. Conceivably, upon choosing the lattice constant of the adsorption lattice suitably, it might be possible to establish phase coherence throughout the whole system, through the tunnelling of individual \paraH2 molecules across adjacent adsorption wells. The ensuing superfluid phase would be similar to the supersolid droplet crystal phase of Refs. \onlinecite{cinti} and \onlinecite{saccani}, with the important conceptual difference that in the present case the adsorption lattice is externally imposed, as opposed to arising from inter-particle interactions.
This  scenario is presently being investigated by computer simulation. 

\section*{Acknowledgments}
This work was supported by the Natural Sciences and Engineering Research Council of Canada (NSERC). Computing support from Westgrid is gratefully acknowledged. MB gratefully acknowledges the hospitality of the Max-Planck Institute for the Physics of Complex Systems in Dresden, and of the Theoretical Physics Institute of the ETH, Z\"urich, where part of this work was carried out.


\begin{thebibliography}{10}
\bibitem{notex}
It is worth mentioning that refiments of the Silvera-Goldman pair potential have been recently proposed, affording a closer reproduction of the experimental equation of state of bulk \paraH2, especially under pressure. See, for instance, Refs. \onlinecite{moraldi} and \onlinecite{tokunbo}.
\bibitem{notesame} It is worth reminding that  in a strict thermodynamic sense one cannot speak of ``phase'' of a finite system, hence the use of adjectives such as ``solidlike" and ``liquidlike", which should be loosely interpreted in terms of qualitative, average properties of the system in its most probable configurations.
\bibitem{note2}
It should be mentioned that the calculation of the superfluid response of  solidlike clusters at low $T$ by means of the ``area" estimator, can sometimes be rendered complicated by the occurrence of rigid rotations of the whole system (or, of one of several concentric rings) in the imaginary time interval $\hbar/k_BT$. Such rotations  give rise to spurious, unphysically large values of the area estimator, masking the genuine superfluid signal. In this study, this was observed to be especially the case for the cluster with $N$=24 molecules.
\end{thebibliography}
\end{document}